\documentclass[doublecolumn]{IEEEtran}
\IEEEoverridecommandlockouts

\usepackage{cuted}
\usepackage{cite}
\usepackage{bm}
\usepackage{amsmath,amssymb,amsfonts}
\usepackage{algorithm}
\usepackage{algorithmic}
\usepackage{subcaption}
\usepackage{diagbox}
\usepackage{caption}
\usepackage{graphicx}
\usepackage{subcaption}  
\usepackage{multirow}
\usepackage{hyperref}
\usepackage{makecell}

\usepackage[letterpaper, left=0.625in, right=0.625in, bottom=1.02in, top=0.70in]{geometry}

\usepackage{textcomp}
\usepackage{xcolor}

\newcommand{\be}{\begin{equation}}
\newcommand{\ee}{\end{equation}}

\def\BibTeX{{\rm B\kern-.05em{\sc i\kern-.025em b}\kern-.08em
    T\kern-.1667em\lower.7ex\hbox{E}\kern-.125emX}}

\begin{document}

\title{Feasibility-Aware Learning-to-Optimize for Wireless Resource Allocation}

\author{Hanwen Zhang \emph{Graduate Student Member, IEEE}, Haijian Sun \emph{Senior Member, IEEE}
\thanks{
H. Zhang and H. Sun (hanwen.zhang@uga.edu, hsun@uga.edu) are with the School of Electrical and Computer Engineering, University of Georgia, Athens, GA, 30602 USA.  \\  Dataset and codes of this paper will be released upon paper acceptance: 
\url{https://github.com/SunLab-UGA/L2O-Constrained-Optimization-Wireless-Comm}}
}

\maketitle

\begin{abstract}
The emergence of 6G wireless communication enables massive edge device access and supports real-time intelligent services such as the Internet of things (IoT) and vehicle-to-everything (V2X). However, the surge in edge devices connectivity renders wireless resource allocation (RA) tasks as large-scale constrained optimization problems, whereas the stringent real-time requirement poses significant computational challenge for traditional algorithms. To address the challenge, feasibility-aware learning-to-optimize (L2O) techniques have recently gained attention. These learning-based methods offer efficient alternatives to conventional solvers by directly learning mappings from system parameters to feasible and near-optimal solutions.  This article provide a comprehensive review of L2O model designs and feasibility enforcement techniques and investigates the application of constrained L2O in wireless RA systems and. The paper also presents a case study to benchmark different L2O approaches in weighted sum rate problem, and concludes by identifying key challenges and future research directions.
\end{abstract}


\section{Introduction}

The emergence of delay-sensitive 6G applications, such as the internet of things (IoT) and vehicle-to-everything (V2X), transforms wireless resource allocation (RA) into high-dimensional constrained optimization problems that require real-time solutions \cite{pivoto2025comprehensive,liu2024survey}. While traditional algorithms struggle with high computational cost, learning-to-optimize (L2O) has emerged as a promising alternative, using neural networks (NNs) to significantly accelerate computing process \cite{donti2021dc}. However, a critical challenge severely hinders the practical deployment of L2O: the lack of feasibility control. Standard NNs may produce solutions that violate the strict operational constraints inherent in wireless systems (e.g., power budgets, rate requirements), potentially compromising system stability and safety. This issue becomes critical in scenarios such as V2X communication, where infeasible solutions compromise system stability and lead to severe safety issues. 

To address this feasibility challenge, a common strategy involves post-processing the NN's raw outputs with traditional projection-based methods, which iteratively map the infeasible solution back into the feasible set.  However, this seemingly straightforward approach introduces significant obstacles for L2O frameworks.  Such iterative procedures often break the differentiability essential for end-to-end training, complicate gradient computation, and are poorly suited for the GPU-accelerated batch processing that underpins modern deep learning.  While some recent works have introduced differentiable convex optimization layers \cite{tang2024pyepo} or differentiable wireless simulation environment \cite{hoydis2022sionna}, they still face scalability issues and high computational overhead in large-scale applications. Therefore, the fundamental tension between feasibility-aware and maintaining an efficient, end-to-end trainable model has become a primary challenge in L2O.

\begin{figure*}[!ht]
    \centering
    \includegraphics[width=6.5in]{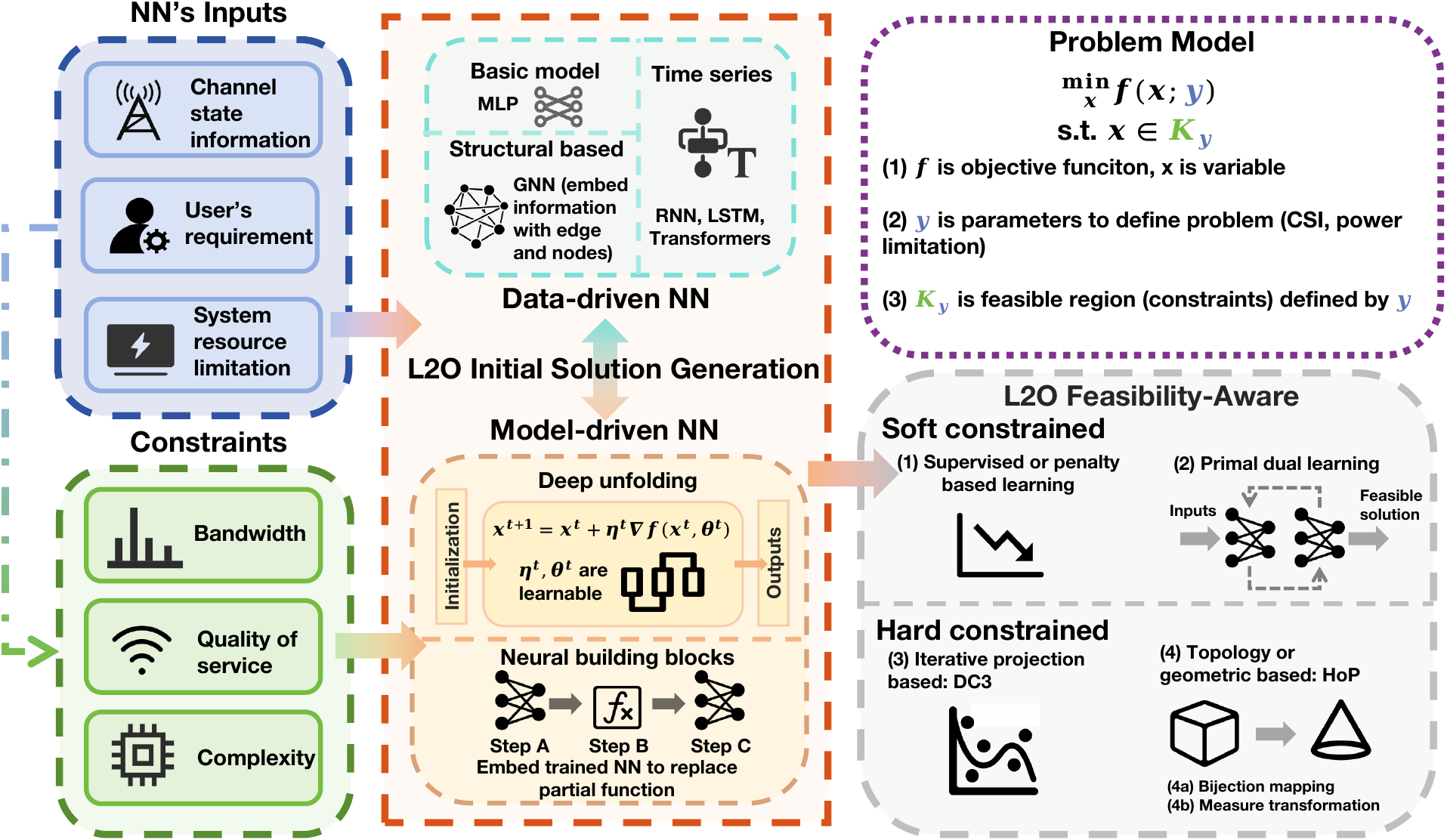}
 \centering
 \caption{The Overview of L2O Structure}
 \label{fig:overview_l2o}
\end{figure*}

Recognizing the limitations of simple penalty and projection methods, the L2O research community has developed more sophisticated techniques for various constraints, offering promising avenues for wireless RA. Early approaches, such as \cite{xu2018semantic}, incorporated constraint violations as penalty terms in the loss function to guide the training process. Hybrid approach, like \cite{park2023self}, combined alternative optimization with learning, where primal variables and Lagrange multipliers are alternately updated through dual NN to mitigate constraint violations. Recent studies have shifted toward L2O methods under hard constraints, which align closely with the high reliability demands in  V2X and IoT. For instance, \cite{donti2021dc,nguyen2025fsnet} introduced a gradient-based correction scheme to explicitly rectify infeasible outputs based on violated constraints. HardNet \cite{min2025hardnethardconstrainedneuralnetworks} and HoP \cite{deng2025hop} further advanced hard constrained L2O by constructing one-to-one differentiable mappings that guarantee feasibility through  enforcement layers and measure transformation, respectively. These approaches offer generalizable and scalable solutions, making them promising candidates for future wireless RA research.

In this article, we focus on leveraging feasibility-aware L2O approaches to address complex constrained optimization problems in wireless RA. We begin by introducing the foundational background of L2O, covering NN architectures and corresponding training strategies. To further illustrate the concept of feasibility aware L2O, we review several representative methods. We then evaluate their effectiveness in practical scenarios through a case study on the quality-of-service (QoS)-aware weighted sum rate (WSR) maximization problem. Finally, we discuss open challenges and outline future research directions in L2O-based wireless RA optimization.

\section{Fundamentals of Learning to Optimize in Constrained Optimization} \label{bg_l2oco}
The application of constrained L2O in wireless RA typically follows a two-phase framework Fig.~\ref{fig:overview_l2o} which includes optimization problem formulation and obtaining feasible solutions. In the past few decades, various wireless RA problems have been formulated, with objective functions ranging from maximizing sum rates or minimizing total power consumption, while the constraints being the bandwidth or individual rate or power requirements. 
As illustrated in Fig.~\ref{fig:overview_l2o}, the overall process is structured as a modular pipeline. The blue module represents the NN inputs, including system parameters such as channel state information, user requirements, and resource limitations. These inputs define both the optimization objective and the feasibility constraints, as shown in the green shaded region. The orange module corresponds to the learnable component of the L2O framework, responsible for generating initial solutions.  {Depending on the availability of domain knowledge, this module may adopt either data-driven or model-driven deep learning. Following initial solution generation, the gray shaded module enforces feasibility by projecting the NN outputs to satisfy system constraints. This part encompasses both soft-constrained and hard-constrained methods which is detailed in the next section. The modularity of this design enables flexible adaptation to a wide range of wireless RA problems, particularly in constraint sensitive scenarios. In this section, we focus on the foundational mechanisms of L2O and the initial solution generation phase which is the orange module in Fig.~\ref{fig:overview_l2o}, including neural architecture design and training strategies. }

\subsection{Overview of L2O Mechanism}\label{Sec:l2o_persp}


\begin{figure}[!ht]
    \centering
    \includegraphics[width=3.3in]{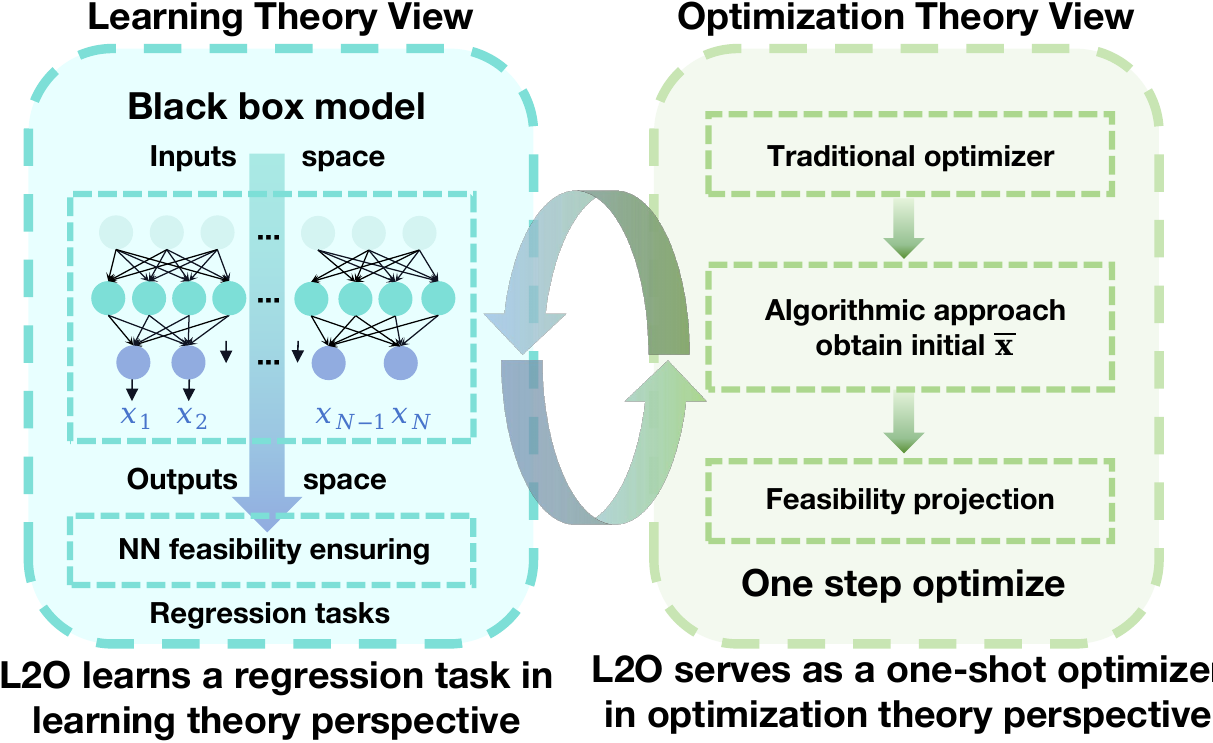}
 \centering
 \caption{{Two Perspectives of Feasibility Aware L2O}}
 \label{fig:perspective_l2o}
\end{figure}
The L2O paradigm presents an efficient, learning-based alternative to traditional iterative algorithms by integrating principles from optimization theory and deep learning. As illustrated in Fig.~\ref{fig:perspective_l2o}, the L2O mechanism can be elucidated from two complementary perspectives. From the optimization perspective, the NN acts as a one-shot parametric optimizer, generating a near-optimal solution in a single forward pass. From a learning theory view, the problem is framed as learning a nonlinear mapping from system parameters to decision variables, which an NN approximates through supervised or self-supervised training. A fundamental guarantee of L2O under either view is that the NN can approximate the underlying input-to-solution mapping. This assumption is theoretically supported by the universal approximation theorem \cite{kim2024minimum}, which states that a NN with sufficient capacity is able to approximate any continuous function on a compact domain. Consequently, if the problem's parameter-to-solution mapping is well-posed (i.e., continuous and unique), the theorem guarantees that an L2O model can asymptotically learn this optimal mapping.

\subsection{NN Designing for Different Application Requirement} \label{Sec:l2o_mec} 
To implement L2O effectively, both the NN architecture and training methodology must be carefully designed. Existing approaches can be broadly classified into two paradigms based on their use of domain knowledge: data-driven learning and model-driven learning. In data-driven methods, optimization is treated as a function approximation problem, where general-purpose architectures learn the input-output mapping purely from data without explicit use of structural priors. In contrast, model-driven approaches embed domain knowledge, included problem structure or algorithmic insights, into the network design to enable model-driven learning.


Data-driven deep learning approaches are particularly effective in scenarios where analytical models are intractable or abundant training data is available. These methods employ general neural architectures to learn implicit input-output mappings directly from datasets. These approaches' primary advantage lies in model-agnostic design where they can approximate complex relationships without requiring explicit modeling of channel conditions or interference structures. Representative data-driven architectures used in L2O frameworks are summarized as follows:

\paragraph{Multilayer Perceptron (MLP)} As a fundamental neural architecture, the MLP consists of stacked fully connected layers, where each layer performs a linear transformation followed by a nonlinear activation such as sigmoid, tanh, or ReLU. Due to its simplicity and ease of implementation, MLPs are widely used in L2O research, serving both as a baseline model and as a flexible module within different L2O frameworks. These properties make MLPs a common choice in general purpose constrained optimization scenarios.

\paragraph{Graph Neural Network (GNN)}
GNNs are designed to process graph-structured data by aggregating information from neighboring nodes. This makes them naturally suited for wireless systems, where user and base station relationships can be modeled as graphs. GNNs offer three key advantages in RA: (1) adaptability to varying network topologies without redesigning the architecture, (2) scalability to large systems via localized message passing, and (3) the ability to embed partial domain-specific structural priors such as channel connectivity or interference patterns. These features make GNNs well-suited for tasks like power control, user association, and spectrum allocation in dynamic wireless environments.

While data-driven methods are effective at learning complex mapping, their black-box behavior and dependence on large training datasets hinder their implementation in practical scenarios requiring interpretability or operating with data scarcity. In contrast, model-driven deep learning integrates prior RA knowledge into NNs' architectures, effectively bridging domain theory and learning capacity. Two primary approaches have emerged to implement this principle systematically:

\paragraph{Deep Unfolding (DU)}
DU systematically embeds learnable parameters into iterative optimization algorithms, enabling accelerated convergence within a fixed number of steps. The process begins by decomposing traditional solvers, such as projection gradient descent or proximal methods, into a predefined number of computational steps. Then these steps are reparameterized by replacing fixed parameters (e.g., step sizes or regularization terms) with layer-specific trainable variables, enabling the model to learn optimal parameters directly from data. The unrolled iterations are subsequently stacked into a periodic architecture, where each layer represents a particular step of the original algorithm. This structured integration of optimization logic into NN design enables DU to simultaneously preserve interpretability and accelerate convergence. These advantages have been demonstrated in various wireless communication applications, including beamforming and channel estimation \cite{shlezinger2023model}. Nevertheless, DU also introduces certain limitations. As the underlying optimization algorithm becomes more complex, especially with higher dimensional inversion, the computational cost of the unrolled NN grows substantially. Furthermore, the architectural design is non-trivial, as the selection of learnable parameters critically impacts both the model's expressiveness and training stability.

\paragraph{Neural Building Blocks}
Unlike algorithm unrolling paradigms, this approach leverages modular neural components to approximate specific mathematical operations within traditional optimization routines (e.g., matrix inversion, Lagrangian multiplier estimation). As illustrated in Fig.~\ref{fig:overview_l2o}, each building block is typically implemented using MLP or GNN architectures. However, unlike purely data-driven models that learn end-to-end mappings over the entire optimization problem, these neural blocks are used to replace specific, well-defined subroutines which preserves the interpretability and modularity of the original solver. Moreover, the framework supports end-to-end or partial training with gradient propagation through both NN and analytical components. This modular design facilitates the construction of hybrid architectures that integrate neural and traditional components, thereby enhancing flexibility and model adaptation during deployment. 
\begin{figure*}[!ht]
    \centering
    \includegraphics[width=6.5in]{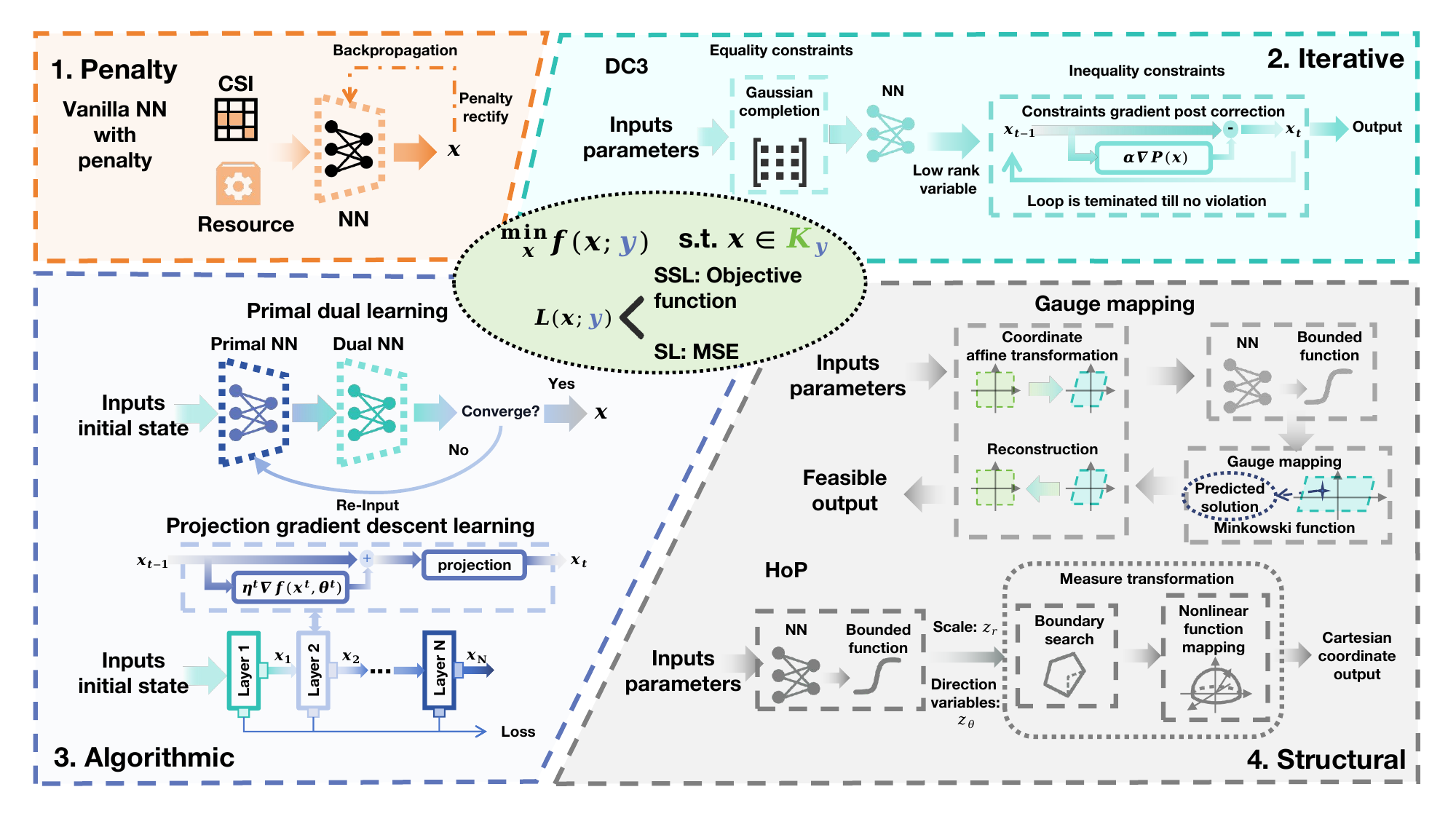}
 \centering
 \caption{Feasibility-Aware L2O Methods}
 \label{fig:methods}
\end{figure*}
\subsection{Specialization for L2O Training Approaches}\label{sec:train_method}
While neural architecture selection plays a pivotal role in implementing specific algorithmic requirements, the training methodology employed in L2O fundamentally determines model generalizability and optimization performance. Current prevalent training paradigms in L2O research are divided into two categories: supervised learning, which relies on ground-truth labels generated by conventional solvers, and label-free methods, such as self-supervised or unsupervised learning.

\paragraph{Supervised Learning (SL)} This approach relies on explicit labels to guide NN training, typically using the mean squared error between network outputs and optimal solutions as the loss function. Early implementations of unconstrained L2O often adopted SL due to its stable convergence, especially in convex settings where smooth gradient landscapes facilitate training \cite{sun2018learning}. However, supervised L2O approaches face two key limitations. First, they do not explicitly capture the underlying relationships between optimization variables and the objective function, which hinders generalization to complex or unseen scenarios. Second, their reliance on large-scale labeled datasets poses practical challenges in wireless communication systems, where generating optimal solutions as supervision labels is often computationally prohibitive. These limitations restrict the scalability and adaptability of supervised L2O in real-world RA tasks.

\paragraph{Self-supervised Learning (SSL) / Unsupervised Learning (UL)} As label-free alternatives, SSL and UL incorporate optimization objectives directly into the loss function. Instead of relying on ground truth labels, these methods use the original objective (e.g., WSR) as the training signal, allowing gradients to be computed from the problem formulation itself. The alignment between the loss function and the problem objective offers two key advantages: it reduces reliance on labeled data and enhances interpretability and also embeds domain-specific priors into the learning process. However, in the absence of supervision, NNs are more susceptible to converging to local minimum or suffering from stagnation. Despite these challenges, SSL and UL remain promising directions for L2O, especially in scenarios where domain knowledge is available while labeled solutions are impractical to obtain.




\section{feasibility-Aware L2O for Constrained Optimization} \label{Sec:feasibility_detail}
Although NNs can theoretically approximate optimal input-output mappings, their practical deployment is often hindered by producing infeasible solutions due to a lack of explicit constraint handling. {To address this issue, feasibility enforcement mechanisms, which are illustrated as the gray module in Fig.~\ref{fig:overview_l2o} and constitute an orthogonal design component independent of neural network architecture choices, are integrated into the L2O framework to ensure that generated outputs remain within the feasible region. These mechanisms can be generally classified into two categories: soft-constrained approaches allow limited constraint violations in exchange for improved objective performance and computational efficiency, which in Fig.~\ref{fig:methods} include penalty-based and algorithmic methods. In contrast, hard-constrained approaches enforce feasibility strictly throughout the learning and inference processes, potentially sacrificing optimality, and are represented by iterative and structural methods in Fig.~\ref{fig:methods}.} This section introduces both approaches along with their implementation details and suitable application scenarios.

\subsection{Soft-constrained L2O}
Soft-constrained L2O provides a flexible learning-based approach that allows minor constraint violations during optimization. Unlike traditional methods that rely on iterative projections or explicit constraint handling, this paradigm incorporates feasibility into the learning process through penalty-augmented loss functions and algorithm-inspired neural architectures. By embedding constraint awareness implicitly into model design, it enables efficient convergence and is suited for scenarios where strict feasibility is not essential.

\paragraph{Penalty-based}
{Penalty-based methods are a widely adopted strategy for soft constraint enforcement in L2O \cite{xu2018semantic}. As shown in penalty part of Fig.~\ref{fig:methods}, these methods incorporate constraint violations into the loss function by adding penalty terms that quantify the degree of infeasibility. This architecture-agnostic approach allows the model to jointly optimize for the objective and constraint satisfaction during backpropagation. While this approach does not guarantee strict feasibility, it reduces computational overhead and preserves compatibility with diverse neural architectures.}

\paragraph{Algorithmic}
Algorithm-informed architectures embed constraint handling directly into the NN by referring to classical optimization procedures. Rather than treating constraints as post-processing steps or handling them indirectly through penalty terms, these methods embed classical optimization principles, such as the Karush-Kuhn-Tucker (KKT) conditions or the alternating direction method of multipliers (ADMM), into the network design to enforce feasibility \cite{park2023self}. A representative example is the primal–dual learning framework, which employs two coordinated networks: a primal network to optimize the original objective and a dual network to update the Lagrangian multipliers that reduce constraint violation.  This approach offers two distinct advantages: First, it establishes a direct correspondence between NN components and optimization variables, enhancing interpretability. Second, it leverages domain knowledge to guide learning, aligning well with the model-driven design principles. However, it also introduces challenges in model training. The coordination between the primal and dual networks is often sensitive to hyperparameter choices such as step sizes and update frequencies. In addition, iterative updates increase memory consumption and may limit scalability in large-scale scenarios




\subsection{Hard-constrained L2O}
Hard-constrained L2O enforces strict feasibility by explicitly projecting the NN's output onto the feasible set, rather than relying on implicit violation control mechanisms such as penalty terms. This projection ensures that all outputs correspond to valid solutions. In practice, equality and inequality constraints are typically treated separately due to their distinct characteristics: equality constraints require the solution to lie exactly on constraint hypersurfaces, while inequality constraints define a half-space for solutions. This separation facilitates more modular projection strategies within the L2O framework.

Building on the definition of hard-constrained L2O, two primary design strategies have been proposed in recent research. The first follows a post-processing paradigm, where the NN may initially produce infeasible outputs, which are then projected onto the feasible set using iterative procedures (e.g., bisection or gradient descent). The second strategy enforces feasibility by construction, designing the output space of the network to be bijectively and differentiably mapped onto the feasible region. This is typically achieved through carefully designed activation functions or measure transformation techniques, ensuring that every output corresponds to a unique and valid solution within the feasible set.
\paragraph{Iterative}
{This class of methods adopts a two-stage strategy: the NN first generates an initial solution, which is subsequently refined through an iterative correction process to ensure feasibility \cite{donti2021dc,nguyen2025fsnet}. A representative example is the DC3 framework \cite{donti2021dc}, which integrates a differentiable correction layer into the L2O pipeline where the iterative part of Fig.~\ref{fig:methods} illustrate the pipline of DC3.} By computing gradients of constraint violations via automatic differentiation, DC3 iteratively performs gradient updates on the NN output until feasibility is achieved. Another recent approach constructs efficient and differentiable enforcement layers \cite{min2025hardnethardconstrainedneuralnetworks}, which guarantees feasibility with input-dependent constraints. 
The key advantage of iterative methods lies in their reliability: with sufficient correction steps, they can consistently produce feasible outputs regardless of the NN's initial estimate. However, since these methods primarily focus on constraint satisfaction, they may compromise performance on the original objective, particularly when the correction path diverges from the optimal trajectory.



\paragraph{Structural}  
{Structural approaches embed constraint geometry directly into the NN architecture, which enables feasibility and avoid iterative post-corrections. These methods construct explicit bijective mappings between estimated output domains and constraint-satisfying solution domains. To clarify the structural approaches pipeline, Fig.~\ref{fig:methods} structural part provides the methodology of gauge mapping and HoP. Gauge mapping ~\cite{li2023learning} which normalizes the output via bounded functions followed by Minkowski projections to maintain geometric alignment with linear constraints. However, such mappings are primarily applicable to bounded linear constraints, such as power or bandwidth. To handle nonlinear or semi-unbounded constraints, recent work \cite{deng2025hop} introduces HoP, a polar coordinate-based transformation, where the NN predicts directional vectors and magnitudes that are mapped back to feasible Cartesian coordinates via differentiable trigonometric functions. This design ensures both bijectivity and differentiability, making it particularly suitable for end-to-end integration in L2O frameworks. The primary strength of structural approaches lies in their inherent satisfaction of constraints without relying on penalty terms or iterative refinements, thus avoiding suboptimality and enabling efficient, fully differentiable learning.}

We have summarized the above L2O approaches and their performance under different metrics in Table~\ref{table:compare}.

\begin{table*}[ht!]
\centering
\caption{Performance Metrics Comparison for Different L2O Approaches}
\begin{tabular}{|l|c|c|c|c|c|} 
\hline
\textbf{Methods / Metrics} &  \textbf{Feasibility Type} &\textbf{Optimality} & \textbf{Constraints Category} & \textbf{Training Difficulty} & \textbf{Inference Speed} \\
\hline
SL \cite{sun2018learning}& Penalty (soft)&High & General  & Easy & Fast \\
\hline
Primal-Dual Learning \cite{park2023self}& Algorithmic (soft)&Mid & General & Mid & Fast \\
\hline
Gauge Mapping \cite{li2023learning} & Structural (hard)&High & Linear Bounded Constraints & Mid & Mid \\
\hline
HardNet \cite{min2025hardnethardconstrainedneuralnetworks}& Structural (hard) &High & Bounded Constraints & Hard & Mid \\
\hline
DC3 \cite{donti2021dc} &Iterative (hard)  &High & General  & Easy & Fast \\
\hline
HoP \cite{deng2025hop} &Structural (hard)  &High & General & Easy & Slow \\
\hline
\end{tabular}
\label{table:compare}
\end{table*}

\section{Numerical Results and Performance Analysis}
This section focuses on applying feasibility-aware L2O approaches to constrained optimization problems in wireless RA. Firstly, we formulate an RA problem based on QoS-aware WSR maximization to benchmark typical L2O approaches. Then, we present a case study comparing the performance of several representative L2O methods under both large-scale and small-scale fading scenarios.

\subsection{RA Problem Formulation}

We consider a downlink multi-user multiple-input single-output system where a base station equipped with 20 antennas serves three single-antenna users. The wireless channels follow a composite fading model, comprising large-scale path loss and small-scale fading modeled by the Nakagami-$m$ distribution. The goal is to maximize the WSR, where each user’s rate depends on the signal-to-interference-plus-noise ratio (SINR), which reflects intra-user interference and additive white Gaussian noise. User priorities are incorporated by assigned weights.

The optimization is subject to two primary constraints: (1) a total power budget that includes both transmission and circuit power consumption, and (2) SINR thresholds to ensure QoS for each user. Due to the non-convex objective and constraints, we employ a traditional WMMSE \cite{zhao2023rethinking} based iterative approach with semidefinite relaxation to derive high-quality feasible solutions that serve as upper-bound baselines.

To enable real-time optimization, we adopt two NN architectures for learning-based solution generation: a standard MLP and a DU model derived from the classical WMMSE algorithm. For constraint enforcement, we explore three methods: a soft-constrained penalty approach, and two hard-constrained projection based frameworks, DC3 \cite{donti2021dc} and HoP \cite{deng2025hop}. Combining these architectures and constraint mechanisms yields  L2O baselines for evaluation.

\subsection{Performance Evaluation}
To comprehensively evaluate the constrained L2O approaches, we conduct two sets of experiments: the first varies the signal-to-noise ratio (SNR) to simulate large-scale fading effects; the second adjusts the Nakagami-$m$ parameter to test robustness under small-scale fading. Each method is evaluated in terms of WSR performance and constraint violation rate. The dataset for each wireless scenario consists of 1,000 samples, which were split into training and testing sets with a 7:3 ratio.

\begin{figure*}[t] 
    \centering

    \begin{subfigure}[b]{0.35\textwidth}
        \centering
        \includegraphics[width=\textwidth]{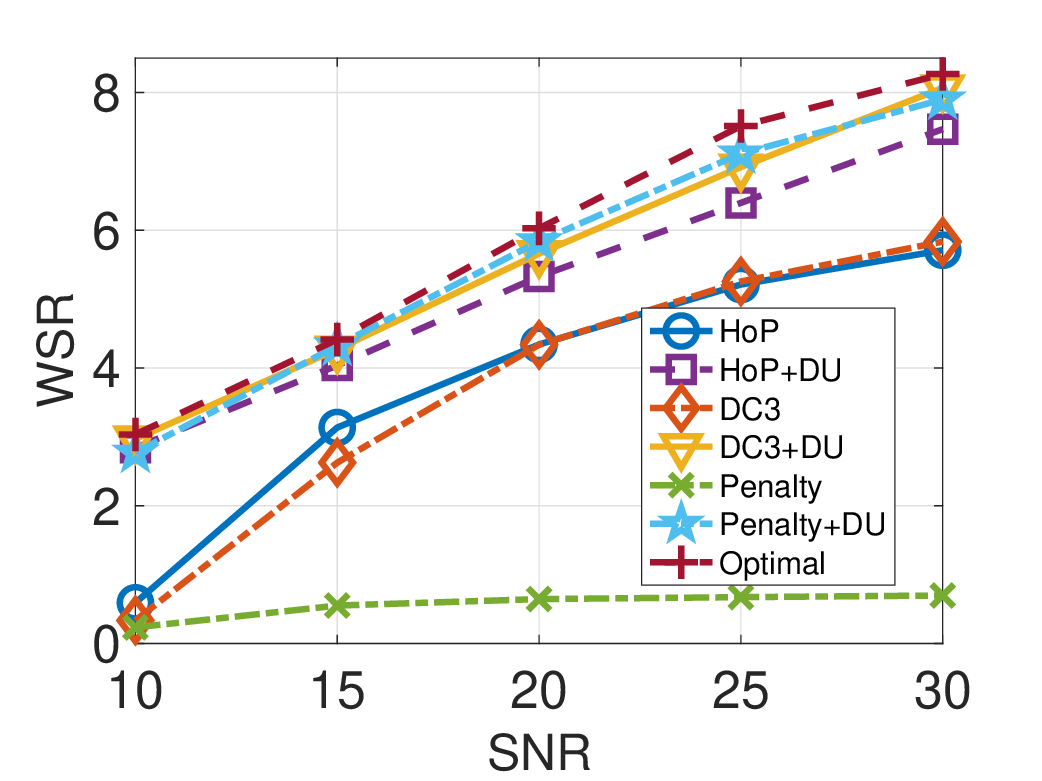}
        \caption{SNR vs. WSR}
        \label{snr_wsr}
    \end{subfigure}
    \hspace{10pt}
    \begin{subfigure}[b]{0.35\textwidth}
        \centering
        \includegraphics[width=\textwidth]{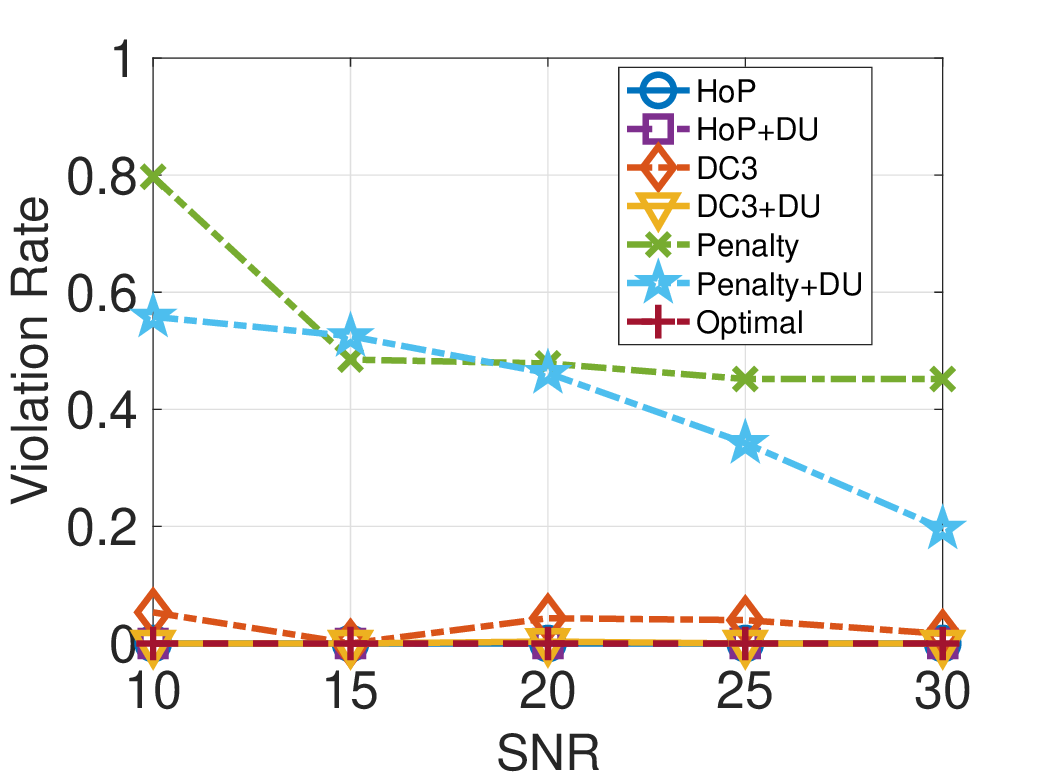}
        \caption{SNR vs. Violation Rate}
        \label{snr_violation}
    \end{subfigure}

    \caption{Constrained L2O Performance over Different Channel SNRs}
    \label{snr_exp}
\end{figure*}

\begin{figure*}[t] 
    \centering

    \begin{subfigure}[b]{0.35\textwidth}
        \centering
        \includegraphics[width=\textwidth]{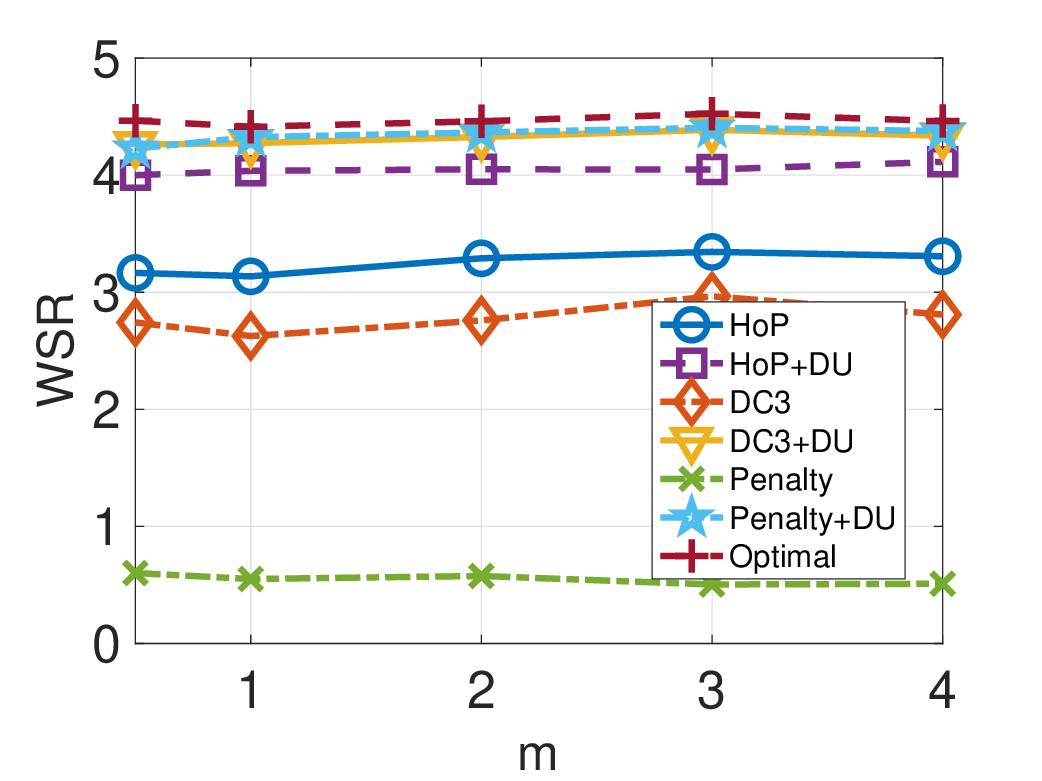}
        \caption{Nagakami-m vs. WSR}
        \label{nagakami_wsr}
    \end{subfigure}
    \hspace{10pt}
    \begin{subfigure}[b]{0.35\textwidth}
        \centering
        \includegraphics[width=\textwidth]{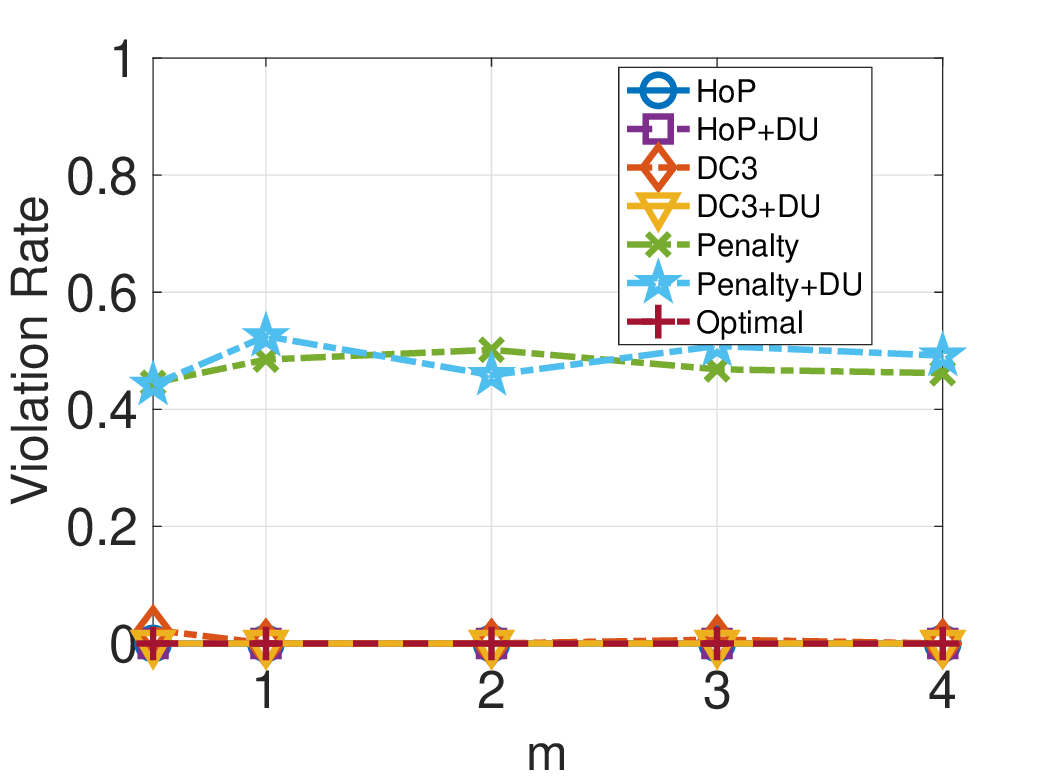}
        \caption{Nagakami-m vs. Violation Rate}
        \label{nagakami_violation}
    \end{subfigure}

    \caption{Constrained L2O Performance over Different Nagakami-m Channel}
    \label{nagakami_exp}
\end{figure*}

From a communication-theoretic perspective, the overall trends of L2O-generated solutions align well with the optimal results given by WMMSE. As shown in Fig. \ref{snr_wsr} and Fig. \ref{nagakami_wsr}, WSR increases rapidly with rising SNR, consistent with improved signal quality. In contrast, variations in the Nakagami-$m$ parameter lead to relatively minor changes in WSR. This can be attributed to the increasing dominance of the line-of-sight path in high-$m$ environments, where stronger interference stabilizes WSR performance.

Furthermore, in the hard constrained L2O approach experiment, the DC3+DU achieves the highest performance which demonstrates the cooperation between a model-driven architecture and a structurally compatible feasibility layer is essential. The DC3+DU stems from the synergy where DU's iterative structure is naturally complemented by DC3's simple, convex-friendly projection. In contrast, HoP, though competitive in standard MLP settings, exhibits reduced performance when combined with DU. This is likely because original HoP outputs both a direction vector and a scale, which complicates its integration with the step-by-step structure of DU, making it less naturally compatible.

Regarding constraint satisfaction (Fig.~\ref{snr_violation} and Fig.~\ref{nagakami_violation}), hard-constrained methods, include DC3 and HoP, achieve near-zero violation rates across all scenarios, reflecting their design guarantees. On the other hand, while penalty-based methods with DU (e.g., penalty+DU) achieve competitive WSR, they still suffer from relatively high violation rates. This confirms that soft constraints reduce computational complexity but cannot ensure feasibility under all conditions.

{Moreover, the experimental results indicate that L2O models achieving both strong performance and low violation rates are those combining model-driven neural architectures with feasibility-aware constraint enforcement. This observation suggests that incorporating optimization structure into NN design inherently facilitates a more favorable balance between optimality and feasibility.}



\section{Future Opportunities and Open Challenges}



While feasibility-aware L2O has been extensively studied in domains such as power systems, operations research, and artificial intelligence, its adoption in wireless communication remains limited where most existing studies in wireless focus on soft-constrained L2O or rely on simple feasibility correction scenarios. This section highlights several open challenges and future research directions for developing constrained L2O methods specific to wireless communication systems.
\paragraph{Expanding the Scope of RA Applications for Constrained L2O}
Wireless communication systems encompass a wide range of constrained optimization problems with diverse and task specific constraints, making feasibility-aware L2O a promising solution framework.   However, most existing studies in this domain remain confined to simplified settings, typically addressing only basic power constraints through penalty terms or projection sub-routines.   In contrast, emerging paradigms such as integrated sensing and communication, reconfigurable intelligent surfaces, and ultra-reliable low-latency communications involve high-dimensional, non-convex, and multi-level constraints that pose substantial challenges to current L2O techniques. Efficiently extending L2O models to handle these various complex constraints will be essential for realizing intelligent and reliable RA in future wireless systems.


\paragraph{Developing Task-Specific L2O Models for Wireless Optimization} 
Unlike generic optimization problems, RA tasks in wireless systems exhibit distinctive structural properties in both objectives and constraints. For example, beamforming problems often involve quadratic objectives and norm-based power constraints. These inherent features provide opportunities to customized task-specific L2O frameworks via the problem constraint geometry. Rather than aiming for general purpose L2O, such designs can exploit domain knowledge of wireless RA problem to improve model performance and interpretability. For instance, when the feasible set is convex and bounded (e.g., $\ell_2$-norm balls in QoS problem), feasibility can be enforced via structured mappings, such as coordinate transformations or adapted activation functions. These task-specific L2O designs not only simplify training but also enhance practicality and performance in deployment.

\paragraph{Multiple Variable Dependency Decoupling}
Most existing L2O approaches are primarily designed for single-variable or decoupled formulations, limiting their effectiveness in capturing the complex interdependencies present in practical wireless systems. In reality, many RA problems, such as multi-user scheduling, beamforming, and coordinated base station control, are inherently multi-variable and highly coupled, where decision variables across users or resources interact in non-separable ways. To accommodate such problems, current methods often reformulate the original formulation into an equivalent single vector representation (e.g., through variable flattening or block-wise stacking).  While this simplification enables training within existing L2O frameworks, it obscures the intrinsic coupling structure and sensitivity among variables. Therefore, developing L2O architectures that natively support multi-variable optimization while considering variable sensitivity and ensuring feasibility, holds great promise for scalable solutions in multi-variables wireless systems.

\paragraph{Embedding Constraints into NN}

Embedding feasibility directly into NN architectures presents a promising direction for constrained L2O, offering the potential to significantly reduce the computational complexity associated with constraint handling. In contrast, most existing hard-constrained L2O frameworks rely on post-correction strategies such as projection based or iterative constraint satisfaction correction, which incur increasing computational overhead as the constraint dimensionality or complexity grows. These procedures often introduce considerable computational overhead which limits the practicality in latency sensitive wireless RA scenarios. Therefore, developing native, structure-aware L2O architectures that inherently enforce feasibility is critical for scalable and real-time optimization in future communication systems.

\section{Conclusions}
This article provides a comprehensive study of feasibility-aware L2O techniques in the context of wireless RA.  We systematically categorized existing L2O approaches based on their constraint-handling strategies, ranging from soft penalty-based methods to hard-constrained formulations using projection and structural embeddings. To benchmark their practical applicability, we conducted experiments on a QoS-aware WSR maximization problem, revealing the superior performance of model-driven, hard-constrained methods with model-driven deep learning in balancing feasibility and optimality. Finally, we provide future opportunities to develop unique feasibility-aware L2O methods for wireless RA problem.


\bibliographystyle{IEEEtran}

\bibliography{lib}

@inproceedings{
donti2021dc,
title={{DC}3: A learning method for optimization with hard constraints},
author={Priya L. Donti and David Rolnick and J Zico Kolter},
booktitle={International Conference on Learning Representations},
year={2021},
url={https://openreview.net/forum?id=V1ZHVxJ6dSS}
}

@misc{min2025hardnethardconstrainedneuralnetworks,
      title={{HardNet}: Hard-Constrained Neural Networks with Universal Approximation Guarantees}, 
      author={Youngjae Min and Navid Azizan},
      year={2025},
      eprint={2410.10807},
      archivePrefix={arXiv},
      primaryClass={cs.LG},
      url={https://arxiv.org/abs/2410.10807}, 
}

@article{deng2025hop,
  title={{HoP}: Homeomorphic Polar Learning for Hard Constrained Optimization},
  author={Deng, Ke and Zhang, Hanwen and Lu, Jin and Sun, Haijian},
  journal={arXiv preprint arXiv:2502.00304},
  year={2025}
}

@inproceedings{xu2018semantic,
  title={A semantic loss function for deep learning with symbolic knowledge},
  author={Xu, Jingyi and Zhang, Zilu and Friedman, Tal and Liang, Yitao and Broeck, Guy},
  booktitle={International Conference on Machine Learning},
  pages={5502--5511},
  year={2018},
  organization={PMLR}
}

@inproceedings{
kim2024minimum,
title={Minimum width for universal approximation using Re{LU} networks on compact domain},
author={Namjun Kim and Chanho Min and Sejun Park},
booktitle={The Twelfth International Conference on Learning Representations},
year={2024},
url={https://openreview.net/forum?id=dpDw5U04SU}
}

@article{tang2024pyepo,
  title={PyEPO: a PyTorch-based end-to-end predict-then-optimize library for linear and integer programming},
  author={Tang, Bo and Khalil, Elias B},
  journal={Mathematical Programming Computation},
  volume={16},
  number={3},
  pages={297--335},
  year={2024},
  publisher={Springer}
}

@article{hoydis2022sionna,
  title={Sionna: An open-source library for next-generation physical layer research},
  author={Hoydis, Jakob and Cammerer, Sebastian and Aoudia, Fay{\c{c}}al Ait and Vem, Avinash and Binder, Nikolaus and Marcus, Guillermo and Keller, Alexander},
  journal={arXiv preprint arXiv:2203.11854},
  year={2022}
}

@article{li2023learning,
  title={Learning to solve optimization problems with hard linear constraints},
  author={Li, Meiyi and Kolouri, Soheil and Mohammadi, Javad},
  journal={IEEE Access},
  volume={11},
  pages={59995--60004},
  year={2023},
  publisher={IEEE}
}

@inproceedings{park2023self,
  title={Self-supervised primal-dual learning for constrained optimization},
  author={Park, Seonho and Van Hentenryck, Pascal},
  booktitle={Proceedings of the AAAI Conference on Artificial Intelligence},
  volume={37},
  number={4},
  pages={4052--4060},
  year={2023}
}

@inproceedings{
nguyen2025fsnet,
title={{FSN}et: Feasibility-Seeking Neural Network for Constrained Optimization with Guarantees},
author={Hoang T. Nguyen and Priya L. Donti},
booktitle={The Thirty-ninth Annual Conference on Neural Information Processing Systems},
year={2025},
url={https://openreview.net/forum?id=oum1txoy1D}
}

@article{pivoto2025comprehensive,
  title={A comprehensive survey of machine learning applied to resource allocation in wireless communications},
  author={Pivoto, Diego Gabriel Soares and de Figueiredo, Felipe AP and Cavdar, Cicek and de Lima Tejerina, Gustavo Rodrigues and Mendes, Luciano Leonel},
  journal={IEEE Commun. Surveys Tuts.},
  year={2025},
  publisher={IEEE}
}

@article{liu2024survey,
  title={A survey of recent advances in optimization methods for wireless communications},
  author={Liu, Ya-Feng and Chang, Tsung-Hui and Hong, Mingyi and Wu, Zheyu and So, Anthony Man-Cho and Jorswieck, Eduard A and Yu, Wei},
  journal={IEEE J. Select. Areas Commun.},
  year={2024},
  month = {Aug.},
  publisher={IEEE}
}

@article{shlezinger2023model,
  title={Model-based deep learning},
  author={Shlezinger, Nir and Whang, Jay and Eldar, Yonina C and Dimakis, Alexandros G},
  journal={Proc. {IEEE}},
  volume={111},
  number={5},
  pages={465--499},
  year={2023},
  month = {Mar.},
  publisher={IEEE}
}

@article{sun2018learning,
  title={Learning to optimize: Training deep neural networks for interference management},
  author={Sun, Haoran and Chen, Xiangyi and Shi, Qingjiang and Hong, Mingyi and Fu, Xiao and Sidiropoulos, Nicholas D},
  journal={IEEE Trans. Signal Processing},
  volume={66},
  number={20},
  pages={5438--5453},
  year={2018},
  month = {Aug.},
  publisher={IEEE}
}

@article{zhao2023rethinking,
  title={Rethinking WMMSE: Can its complexity scale linearly with the number of BS antennas?},
  author={Zhao, Xiaotong and Lu, Siyuan and Shi, Qingjiang and Luo, Zhi-Quan},
  journal={IEEE Trans. Signal Processing},
  volume={71},
  pages={433--446},
  year={2023},
  publisher={IEEE}
}

\end{document}